\documentclass[conference]{IEEEtran}
\IEEEoverridecommandlockouts
\usepackage{cite}
\usepackage{amsmath,amssymb,amsfonts}
\usepackage{graphicx}
\usepackage{textcomp}
\usepackage{xcolor}
\usepackage[colorlinks=true, allcolors=blue]{hyperref}
\usepackage{booktabs}
\usepackage{tabularx}
\usepackage{enumitem}
\usepackage{float}

\begin{document}

\title{Edge-Deployable LLM Fine-Tuning on a Single GPU for Telecom Network Troubleshooting}

\author{\IEEEauthorblockN{Chenhua Shi\textsuperscript{*}, Bhavika Jalli\textsuperscript{*}, John Zou\textsuperscript{*}, Gregor Macdonald, Wanlu Lei, Mridul Jain, Joji Philip}
\IEEEauthorblockA{Ericsson\\
\textsuperscript{*}Equal contribution}}
\maketitle

\begin{abstract}
Telecom troubleshooting at edge sites requires low-latency model responses that cannot tolerate cloud round-trip delays, while data sovereignty requirements necessitate localized model adaptation; however, deploying GPUs at cell sites is fundamentally constrained by power, cooling, space, and weight limitations, and further challenged by RAN traffic patterns that lead to low hardware utilization and poor return on investment, as well as architectural mismatches between ASIC-optimized deterministic processing and GPU-based AI workloads. Consequently, single-GPU fine-tuning emerges as a critical enabler for practical AI deployment at the edge rather than merely a resource constraint. In this paper, we present a GPU profiling study of LLM fine-tuning under the Unsloth framework, systematically characterizing the impact of maximum sequence length, GPU memory utilization, Low-Rank Adaptation (LoRA) rank, and number of generations on a single edge-class accelerator. We analyze trade-offs between parameter configurations and KV cache usage, examine the effect of inductor compilation on runtime stability, and quantify activation memory overhead during training. Furthermore, we demonstrate that model characteristics—specifically reasoning versus non-reasoning architectures—significantly influence both supervised fine-tuning (SFT) and reinforcement fine-tuning (RFT) due to differences in chat template requirements, including reasoning tags and control flags. Experiments are conducted on a telecom troubleshooting dataset comprising question–answer pairs augmented with top-3 retrieved context documents. The results provide practical configuration guidelines for enabling efficient, stable, and resource-aware LLM fine-tuning in telecom edge environments.
\end{abstract}


\section{Introduction}

Telecom operators increasingly deploy AI-assisted troubleshooting at edge sites, including regional Network Operations Centers (NOCs), O-RAN near-Real-Time RAN Intelligent Controller (near-RT RIC) nodes~\cite{oran_aiml}, and ETSI Multi-access Edge Computing (MEC) platforms~\cite{etsi_mec003}, where compute is deliberately constrained to a single GPU accelerator. At these sites, large language models (LLMs) must be adapted to operator-specific fault patterns, alarm vocabularies, and network topologies. However, offloading fine-tuning to centralized cloud introduces round-trip latencies incompatible with real-time network troubleshooting, and transmitting sensitive operator data (subscriber metadata, performance counters, alarm logs) to remote infrastructure violates data sovereignty requirements imposed by regulations such as GDPR. Single-GPU fine-tuning at the edge is therefore not merely a resource constraint but a deployment requirement. The NVIDIA RTX A6000, representative of accelerators available in edge-class servers and regional NOCs~\cite{nvidia_egx}, provides sufficient capacity for parameter-efficient adaptation of 7--8B parameter models using Low-Rank Adaptation (LoRA)~\cite{lora2021} and quantization, yet the operational limits of such hardware for LLM fine-tuning remain poorly characterized.

This work addresses that gap by characterizing LLM fine-tuning using the Unsloth framework~\cite{unsloth_guide} on a single edge-class GPU. Our study pursues two objectives: (1) GPU memory profiling to establish safe operating envelopes for edge-deployed accelerators, and (2) identifying behavioral differences between reasoning and non-reasoning models that affect edge deployment choices. We fine-tune the Qwen family of models on a telecom troubleshooting dataset comprising question-answer pairs with top-3 retrieved document chunks, profiling memory consumption, batch size limits, LoRA rank, and maximum sequence length within a single GPU budget. Through supervised fine-tuning (SFT) followed by reinforcement fine-tuning (RFT), we demonstrate that edge-local adaptation yields production-quality telecom models without cloud infrastructure.

The main contributions of this paper are as follows:
\begin{itemize}[leftmargin=*]
\item Enable edge-deployable LLM fine-tuning for telecom by demonstrating that domain-adapted models can be trained entirely on a single edge-class GPU such as RTX A6000, providing a practical workflow validated on a telecom troubleshooting dataset.
\item Conduct a systematic GPU profiling study under the Unsloth framework, quantifying trade-offs between sequence length, KV cache usage, and GPU memory utilization to establish safe operating envelopes for edge-deployed accelerators.
\item Identify and analyze behavioral inconsistencies between reasoning and non-reasoning models, including a hybrid tokenizer strategy that enables reasoning-aware training, highlighting the need for model-specific chat template configurations at the edge.
\end{itemize}

Our experiments validate the proposed workflow on the telecom troubleshooting dataset in an edge-representative single-GPU setting. We observe that DeepSeek-R1, when fine-tuned using the hybrid tokenizer strategy, achieves substantial gains in both factual grounding and response consistency compared to Qwen2.5-7B. These results demonstrate that careful profiling, template alignment, and single-GPU optimization at edge sites can yield domain-adapted models with improved reasoning faithfulness and resource efficiency, without requiring centralized cloud training infrastructure.

\section{Related Work}

\textbf{LLM Fine-Tuning for Specialized Domains.} Decoder-only Transformers have demonstrated broad generalization across NLP tasks, but pre-training remains prohibitively resource-intensive~\cite{brown2020gpt3}. Fine-tuning aligns general LLMs to domain semantics at substantially lower cost. Parameter-efficient approaches such as LoRA and QLoRA reduce trainable state and VRAM by inserting low-rank adapters and quantizing frozen weights, enabling single-GPU adaptation~\cite{lora2021,dettmers2023qlora}. Recent work documents how LLMs assist telecom workflows (alarm correlation, root cause analysis) and why domain-grounded retrieval is essential to control hallucinations.

\textbf{Memory Management and KV Cache.} At long contexts, the KV cache dominates memory and shapes batching decisions~\cite{nvidia_kv_cache_2023}. vLLM's PagedAttention mitigates fragmentation~\cite{kwon2023vllm}, while KV cache quantization extends practical context windows~\cite{hooper2024kvquant}. Serving studies such as DistServe~\cite{zhong2024distserve} and SARATHI~\cite{agrawal2023sarathi} clarify prefill/decode bottlenecks that inform training-time generation for RFT.

\textbf{Reinforcement Post-Training and Tooling.} GRPO reduces critic memory via group-relative baselines and strengthens reasoning behaviors~\cite{deepseekmath}; implementations are available in TRL~\cite{trl_grpo}. Chat templates strongly affect reasoning visibility: Hugging Face's \texttt{apply\_chat\_template} exposes controls such as \texttt{add\_generation\_prompt}, while Qwen variants introduce thinking switches~\cite{hf_chat_templates,qwen3_docs}. Unsloth integrates PEFT/quantization with fast kernels for single-GPU fine-tuning~\cite{unsloth_guide}, and PyTorch~2's \texttt{torch.compile} shifts memory/time trade-offs between compilation and steady-state execution.

\textbf{Edge AI for Telecom.} The O-RAN near-RT RIC architecture hosts AI/ML inference workloads for network optimization~\cite{oran_aiml}, and ETSI MEC specifications now address on-site AI/ML lifecycle management including model training~\cite{etsi_mec031}. However, the majority of prior work on edge intelligence focuses on deploying pre-trained models for inference at the edge rather than training or fine-tuning models locally~\cite{edge_ai_telecom}. Federated learning approaches~\cite{fl_telecom} address distributed training across multiple sites but do not characterize the single-site GPU profiling needed for practical deployment at individual edge nodes. A gap therefore remains in understanding the resource envelopes and configuration trade-offs for LLM fine-tuning on the single-GPU accelerators typically available at telecom edge sites.

\textbf{Positioning.} Closest to our study, Xia et al.~\cite{xia2024singlegpu} provide a single-GPU fine-tuning characterization with MoE emphasis. Our work extends this by: (i) characterizing dense models under long contexts on 48\,GB VRAM; (ii) quantifying joint effects of sequence length, KV-cache utilization, LoRA rank, and generation count; (iii) measuring template-driven behavioral differences between reasoning and non-reasoning models on a telecom dataset; (iv) providing the first GPU profiling characterization specifically targeting edge deployment for telecom, establishing safe operating envelopes for single-GPU fine-tuning at edge sites.

\section{Methodology}

To fine-tune a telecom-specific model capable of generating detailed stepwise solution plans for troubleshooting network faults based on ingested documents within a knowledge graph, we adopt a two-stage approach designed for data-scarce edge environments where expert-labeled telecom data is limited and expensive to collect at individual sites. We begin with SFT as a warm-up phase, where the model is trained offline to replicate ground-truth outputs from a static dataset. In this stage, the model learns structural formats such as step-by-step plans, the use of \texttt{<reasoning>} tags in non-reasoning models, and \texttt{<answer>} tags across both reasoning and non-reasoning models. Following SFT, we employ RFT, which is particularly effective under limited data availability. By training SFT on a small expert-curated seed set and then applying RFT on synthetically generated data, the pipeline is practical for edge sites where assembling large labeled corpora from local fault logs is prohibitively costly. RFT has been shown to outperform SFT in data-scarce regimes while enhancing reasoning capabilities through chain-of-thought alignment~\cite{predibase_rft}. RFT guides the model's behavior through feedback and rewards instead of direct labels. Specifically, we leverage Group Relative Preference Optimization (GRPO), an efficient and widely adopted strategy that reduces GPU memory overhead, as demonstrated in prior work such as DeepSeek-R1-Zero~\cite{deepseekmath}.

During RFT, we utilize the Unsloth framework on a single GPU, which integrates quantization, FlashAttention2, XFormers, and KV cache acceleration. To ensure model quality without cloud-scale validation infrastructure, we design reward functions addressing format correctness (regex-based tag validation), quality control (penalizing repetitive or verbose responses), reasoning grounding (verifying reasoning traces against QA pairs), and domain-specific criteria (chronological troubleshooting steps with relevant performance counters). We adapt RAGAS~\cite{ragas} to score faithfulness, relevance, and correctness with respect to retrieved context. Both rule checks and RAGAS scoring run locally on the same edge GPU, eliminating the need for external validation services.

Finally, we systematically vary parameters including maximum sequence length, LoRA rank/alpha, target modules, GPU memory utilization, training epochs, batch size, and number of generations to determine the operational envelope for edge-deployed GPUs with fixed VRAM budgets. Understanding these limits is essential for automated edge training pipelines where models must be retrained on-site without manual intervention.

\section{Experiments}

We conduct our experiments on an NVIDIA RTX A6000 GPU with 48 GB of VRAM, representative of the single-GPU accelerators deployed at regional NOCs and edge aggregation points~\cite{nvidia_egx}. For GPU memory profiling and optimization, we fine-tune the unsloth/Qwen2.5-7B model~\cite{qwen2.5} under different combinations of configuration settings. In addition, we employ unsloth/DeepSeek-R1-0528-Qwen3-8B-unsloth-bnb-4bit~\cite{deepseekr1} as a reasoning-oriented model to investigate behavioral differences between reasoning and non-reasoning LLMs, particularly with respect to chat template requirements and flag settings.

\subsection{Experimental Setup}

\textbf{Models.} We fine-tune two pre-trained models: Qwen2.5-7B (non-reasoning) and DeepSeek-R1-0528-Qwen3 (reasoning), following a two-stage process of SFT followed by RFT. In particular, DeepSeek-R1-0528-Qwen3 is obtained by distilling the chain-of-thought from DeepSeek-R1-0528 into the Qwen3-8B Base model through post-training. Qwen2.5-7B has 7.61B parameters with 28 layers and attention heads configured as 28 (Q) and 4 (KV) using grouped-query attention. DeepSeek-R1-0528-Qwen3-8B has 8.2B parameters with 36 layers and attention heads of 32 (Q) and 8 (KV). Both models support context lengths up to 131,072 tokens. Notably, both models' parameter counts (7--8B) fall within the capacity of edge-class single-GPU accelerators when combined with 4-bit quantization, making them practical choices for on-site fine-tuning at telecom edge nodes. Different chat templates are applied for the two models in both SFT and RFT to account for their distinct behaviors. For parameter-efficient fine-tuning (PEFT), we adopt LoRA and restrict updates to the \texttt{q\_proj}, \texttt{k\_proj}, and \texttt{v\_proj} modules in order to remain within the single-GPU memory budget.

\textbf{Datasets.} We employ a network troubleshooting dataset~\cite{shi2025thinkless}, representative of operational data that must be trained locally at edge sites due to data sovereignty and latency requirements. Questions cover alarm diagnosis and fault resolution; answers provide step-by-step solution plans integrating alarms, configuration data, and performance counters. The SFT dataset augments each QA pair with top-3 retrieved document chunks and a reasoning trace; the RFT dataset includes only the top-3 chunks. The dataset comprises 50 SME-validated seed pairs and 500 synthetically generated RFT pairs. Table~\ref{tab:qa_example} illustrates a representative question-answer pair.

\begin{table}[t]
\centering
\caption{Example Question-Answer Pair from the RFT Dataset}
\label{tab:qa_example}
\scriptsize
\begin{tabularx}{\linewidth}{p{0.22\linewidth}X}
\toprule
\textbf{Question} & How would you troubleshoot and resolve an issue where the PowerDistributionTemperature is too high, leading to the High Operating Temperature Main Load Disconnected alarm? \\
\midrule
\textbf{Answer} & 
\textbf{1.} Check for correlated alarms (No Connection, Fan Failure, Fan Power Supply Failure, Clogged Air Filter) raised by FanGroup, FieldReplaceableUnit, or Cabinet.
\textbf{2.} Examine site conditions: verify cooling capacity, unobstructed airflow, and ambient temperatures are within range.
\textbf{3.} Inspect climate system: check fans, coolers, heaters; verify thermosiphon systems and heat exchangers for outdoor cabinets.
\textbf{4.} Check RRU-specific conditions: ensure free airflow, verify solar shield is intact, confirm nearby equipment is not overheating the unit.
\textbf{5.} Restart the faulty unit using \texttt{restartRank RESTART\_COLD\_W\_TEST} with \texttt{RestartReason} set to \texttt{UNPLANNED\_COLD\_WITH\_HW\_TEST}.
\textbf{6.} Consult maintenance support if the alarm persists after all steps. \\
\bottomrule
\end{tabularx}
\end{table}

\textbf{GPU Profiling Setup.} To establish the operational envelope for edge-deployed hardware, we profile GPU resource consumption across training configurations. Since each QA pair includes top-3 retrieved document chunks, effective context lengths are large. We apply context-weighted optimization: a maximum context length is set per chunk with higher weights for top-ranked documents, extracting the most important sections and dynamically reallocating remaining budget to subsequent chunks. The average token length for the top-3 chunks is approximately 16,126.

We then profile GPU performance by varying key parameters before and during training, including maximum sequence length, GPU memory utilization, LoRA rank, and number of generations. The experiments are run on a 4-bit Qwen2.5-7B model that has 3.805 GB model weights. For RFT, we begin with two generations per prompt, a LoRA rank of 8, and GPU utilization of 0.7. The maximum sequence length is gradually increased from 1k to 80k tokens. Out-of-memory errors (CUDA illegal memory access) occur at 60k and 80k tokens, so we fix the sequence length at 50k and vary GPU utilization from 0.3 to 0.9. Additionally, we increase the number of generations to 4 and 6 while holding GPU utilization at 0.7.

\subsection{Analysis of GPU Profiling}

From our GPU profiling experiments, several key findings emerge. Table~\ref{tab:GPUBeforeTraining} records GPU memory across configurations prior to training. Increasing maximum sequence length raises compilation overhead while reducing KV cache token capacity. GPU memory utilization governs KV cache allocation: higher values expand cache but reduce memory available for activation weights. Generations have little effect on memory, though they extend training time. Sequence lengths above 60k consistently triggered CannotAccessIllegalMemory (CAIM) errors, while low GPU utilization with large sequence lengths caused Cache Block Errors from insufficient cache memory after compilation overhead.

\begin{figure}[t]
\centering
\includegraphics[width=\linewidth]{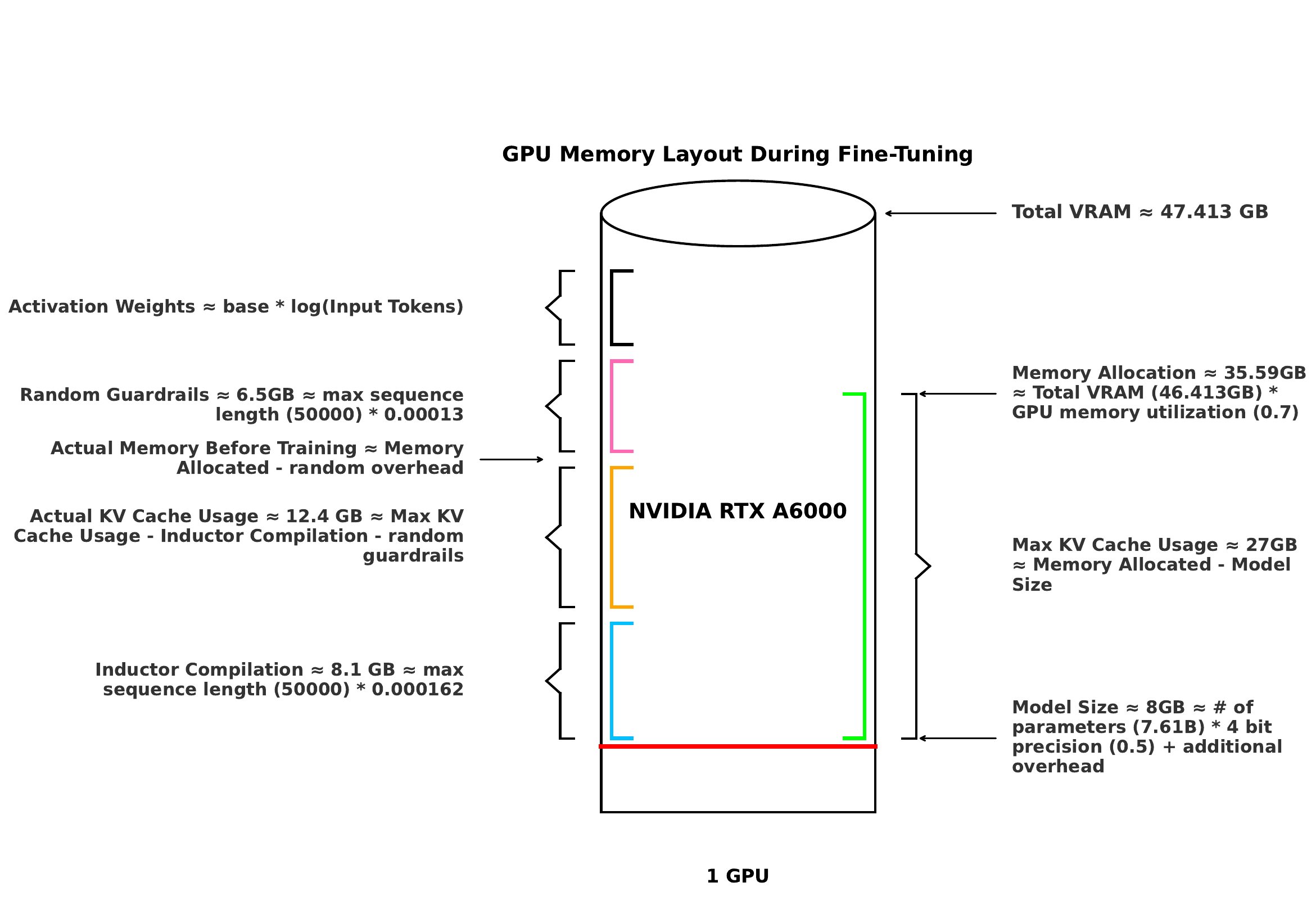}
\caption{GPU memory usage prior to training on an NVIDIA RTX A6000 GPU, configured with max sequence length of 50,000 and GPU utilization set to 0.7.}
\label{fig:GPUProfiling}
\end{figure}

The trade-off is clear: longer sequences reduce KV cache headroom, while higher GPU utilization constrains activation weights. We recommend setting maximum sequence length to the 75\% quantile of dataset requirements and tuning GPU utilization just high enough for sufficient KV cache. For edge operators, 50k tokens with 0.7 utilization represents the safe operating point on 48\,GB accelerators. Figure~\ref{fig:GPUProfiling} illustrates GPU memory usage at this configuration.

\begin{table}[t]
\centering
\caption{GPU Profiling Results Before Training (Qwen2.5-7B, 4-bit)}
\label{tab:GPUBeforeTraining}
\scriptsize
\begin{tabular}{cccccl}
\toprule
\textbf{SeqLen} & \textbf{Util} & \textbf{Compile} & \textbf{KV Cache} & \textbf{Tokens} & \textbf{Result} \\
\midrule
1k & 0.7 & 8.3 GB & 24.9 GB & 438k & Success \\
5k & 0.7 & 8.8 GB & 24.5 GB & 442k & Success \\
10k & 0.7 & 9.6 GB & 23.6 GB & 437k & Success \\
20k & 0.7 & 11.5 GB & 20.4 GB & 415k & Success \\
40k & 0.7 & 14.6 GB & 14.3 GB & 358k & Success \\
50k & 0.7 & 16.2 GB & 11.2 GB & 329k & Success \\
60k & 0.7 & 17.3 GB & -- & -- & CAIM Error \\
80k & 0.7 & 20.4 GB & -- & -- & CAIM Error \\
\midrule
50k & 0.3 & 16.2 GB & -- & -- & Cache Error \\
50k & 0.5 & 16.2 GB & 1.6 GB & 153k & Success \\
50k & 0.6 & 16.2 GB & 6.4 GB & 241k & Success \\
50k & 0.8 & 16.2 GB & 15.7 GB & 411k & Success \\
50k & 0.9 & 16.2 GB & 15.7 GB & 411k & Success \\
\midrule
50k & 0.7 & 16.2 GB & 11.2 GB & 329k & Success (4 gens) \\
50k & 0.7 & 16.2 GB & 11.2 GB & 329k & Success (6 gens) \\
\bottomrule
\end{tabular}
\end{table}

From Table~\ref{tab:GPUDuringTraining}, when input tokens exceed 10k with 0.7 GPU utilization and 50k sequence length, training consistently encounters OOM errors. This occurs because activation weight requirements grow with input tokens, eventually exceeding VRAM capacity. These OOM boundaries are critical for automated edge pipelines, since an OOM at step~155 of 250 wastes hours of compute. At 5k input tokens, training completes in under 3~hours, enabling daily model refresh cycles within overnight maintenance windows.

\begin{table}[t]
\centering
\caption{GPU Profiling Results During Training (Qwen2.5-7B, 4-bit)}
\label{tab:GPUDuringTraining}
\scriptsize
\begin{tabular}{cccccl}
\toprule
\textbf{Input} & \textbf{SeqLen} & \textbf{Util} & \textbf{Gens} & \textbf{Steps} & \textbf{Duration/Outcome} \\
\midrule
1k & 50k & 0.7 & 2 & 250 & 116 min \\
5k & 50k & 0.7 & 2 & 250 & 173 min \\
10k & 50k & 0.7 & 2 & 250 & 229 min \\
20k & 50k & 0.7 & 2 & 250 & OOM @ step 155 \\
20k & 50k & 0.6 & 2 & 250 & 323 min \\
20k & 40k & 0.7 & 2 & 250 & OOM @ step 20 \\
20k & 40k & 0.6 & 2 & 250 & 312 min \\
5k & 50k & 0.7 & 4 & 250 & 241 min \\
5k & 50k & 0.7 & 6 & 250 & 289 min \\
10k & 50k & 0.7 & 4 & 250 & OOM @ step 0 \\
10k & 50k & 0.7 & 6 & 250 & OOM @ step 0 \\
10k & 50k & 0.6 & 4 & 250 & 313 min \\
\bottomrule
\end{tabular}
\end{table}

\subsection{Cross-Architecture Validation}

To assess generalizability, we replicated key experiments on Llama-3.1-8B-Instruct~\cite{lama3.3} using identical hardware and dataset configurations.

\textbf{Compilation Behavior.} The 60k compilation cliff in Qwen2.5-7B does \textit{not} occur in Llama-3.1-8B, which trains successfully at 60k sequences. This stems from architectural differences: Qwen uses 28 Q-heads and 4 KV-heads, while Llama employs 32 Q-heads and 8 KV-heads, resulting in different memory allocation patterns that avoid CAIM-triggering fragmentation. For edge deployment, operators must profile per-architecture safe limits before deploying automated training; configurations validated on one model family cannot be assumed safe for another.

\textbf{Batch Size Constraints.} The batch size constraint generalizes across architectures: both models encounter OOM when batch size $\geq 4$ on sequences exceeding 20k tokens, indicating a fundamental VRAM limitation tied to activation memory scaling.

\textbf{Training Duration.} Llama-3.1-8B completes 250 steps in $\sim$29 minutes at 20k input tokens, significantly faster than Qwen2.5-7B's 229 minutes at 10k tokens. For edge sites with tight retraining windows, Llama's faster training enables more frequent model updates.

Table~\ref{tab:cross_arch} summarizes these results, underscoring that sequence length thresholds and compilation behaviors are not portable across model families.

\begin{table}[t]
\centering
\caption{Cross-Architecture Validation (Llama-3.1-8B-Instruct)}
\label{tab:cross_arch}
\scriptsize
\begin{tabular}{ccccl}
\toprule
\textbf{Input} & \textbf{SeqLen} & \textbf{Util} & \textbf{Gens} & \textbf{Outcome} \\
\midrule
20k & 50k & 0.7 & 2 & 29 min (Success) \\
20k & 60k & 0.7 & 2 & 29 min (Success) \\
20k & 50k & 0.7 & 4 & OOM @ step 17 \\
\bottomrule
\end{tabular}
\end{table}

\subsection{Analysis of Model Behavioral Differences}

Unsloth's FastLanguageModel and HuggingFace's AutoTokenizer serve distinct roles: the former handles memory-efficient model loading, while the latter manages tokenization and chat template application. During fine-tuning, we observe that the Unsloth DeepSeek-R1-Qwen-3-8B chat template automatically strips \texttt{<think>} tag content during \texttt{apply\_chat\_template}, preventing reasoning traces from reaching the model during SFT. In contrast, the \texttt{AutoTokenizer} for Qwen-3-8B preserves \texttt{<think>} tags during SFT but does not surface them at inference time. Conversely, DeepSeek-R1-Qwen-3-8B automatically inserts \texttt{<think>} tags at inference, making reasoning visible. We also find that combining \texttt{<custom\_reasoning>} with \texttt{<think>} tags causes redundant double reasoning and should be avoided, and that \texttt{enable\_thinking} is only effective via \texttt{AutoTokenizer} when \texttt{add\_generation\_prompt=True}.

From these findings, we derive the following solution strategy:
\begin{itemize}[leftmargin=*]
\item Use the \texttt{AutoTokenizer} (Qwen-3-8B) chat template for SFT to preserve reasoning traces during training, with \texttt{add\_generation\_prompt=True} and \texttt{enable\_thinking=False}.
\item Use Unsloth DeepSeek-R1-Qwen-3-8B for inference (via callback trainer) and RFT, with the same configuration to leverage its built-in reasoning visibility.
\end{itemize}

This hybrid strategy leverages the complementary strengths of both tokenizers, enabling reasoning-aware training while ensuring consistent and interpretable reasoning outputs during inference.

\subsection{Edge Deployment Considerations}

\textbf{Inference Latency.} With models fine-tuned locally, inference runs on the same edge GPU, eliminating cloud round-trip latency. The 7--8B parameter models serve requests in seconds on the RTX A6000, meeting responsiveness requirements at regional NOCs and near-RT RIC nodes.

\textbf{Model Update Cadence.} At 5k input tokens, training completes in $\sim$173 minutes ($\sim$3~hours), enabling overnight retraining. At 10k tokens, $\sim$229 minutes ($\sim$4~hours) remains feasible for daily updates, allowing edge sites to incorporate new fault patterns within 24~hours.

\textbf{Data Sovereignty.} Edge-local training keeps sensitive operator data (alarm logs, performance counters, subscriber metadata) on-site throughout the pipeline, satisfying GDPR and operator security policies without data anonymization or cloud transfer, aligning with the ETSI MEC framework~\cite{etsi_mec003}.

\textbf{Hardware Selection.} Llama-3.1-8B handles 60k sequences where Qwen encounters CAIM errors, with 7$\times$ faster training (29 vs.\ 229 minutes), making it attractive for time-constrained edge sites. However, Qwen's stronger evaluation metrics (Table~\ref{tab:model_means}) may justify longer training for quality-critical applications. Operators must balance training efficiency against model quality.

\subsection{Proposed Workflow for Single-GPU Fine-tuning}

Building on the characterization study, we propose the following practical workflow for deploying LLM fine-tuning with Unsloth on a single GPU at telecom edge sites:

\begin{enumerate}[leftmargin=*]
\item \textbf{Model selection.} Select a model constrained by the edge site's VRAM budget (48\,GB). Prefer 7--8B parameter models with 4-bit quantization and verify chat template and context window support.
\item \textbf{SFT dataset curation.} Curate from local fault logs and operator knowledge bases. Include reasoning traces for non-reasoning models; omit for reasoning models when templates strip \texttt{<think>} content.
\item \textbf{SFT configuration.} Verify chat template behavior (tag preservation vs.\ stripping). Set \texttt{max\_seq\_length} to $\sim$75\% quantile of observed contexts. Configure LoRA rank/alpha within the edge GPU's memory envelope.
\item \textbf{Load SFT adapters into RFT model.} Ensure model type, parameter count, and quantization match. Validate checkpoint compatibility for automated pipeline handoff.
\item \textbf{RFT configuration.} Start conservative (shorter inputs, moderate GPU utilization, small LoRA rank) for unattended training. Scale up only after profiling confirms stability.
\item \textbf{GPU profiling \& tuning.} Treat as edge site commissioning: profile memory at load, compile, cache allocation, and training. Record per-architecture limits (e.g., the 60k compilation cliff for Qwen) as deployment constraints.
\item \textbf{Context management.} Compress long contexts via weighted pruning when input tokens exceed 10k to stay within the VRAM budget for stable unattended training.
\item \textbf{Evaluate and iterate.} Use locally-runnable RAGAS metrics on the same edge GPU. Iterate hyperparameters targeting overnight retraining windows.
\end{enumerate}

\subsection{Experimental Results}

\begin{figure}[t]
\centering
\includegraphics[width=\linewidth]{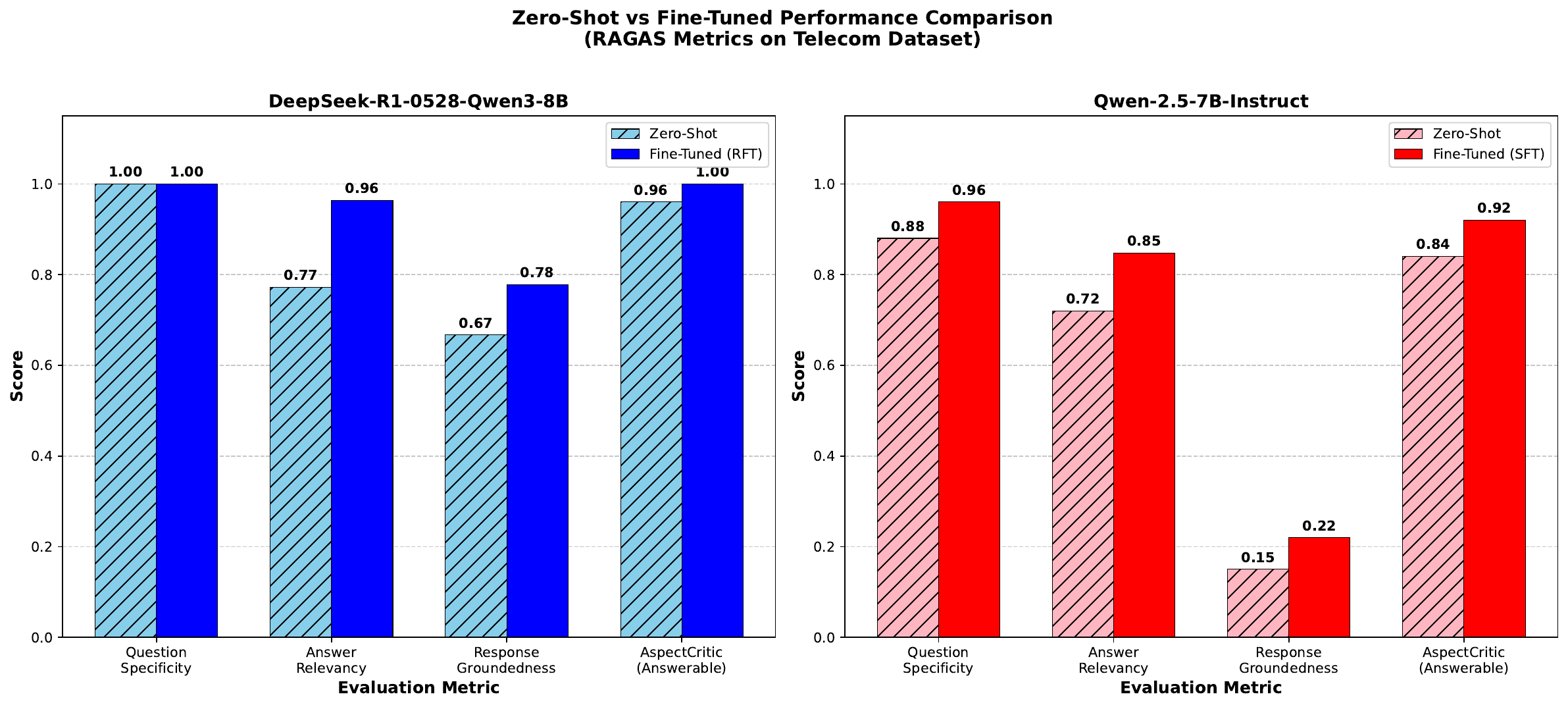}
\caption{Zero-shot vs.\ fine-tuned performance comparison on RAGAS evaluation metrics for DeepSeek-R1 and Qwen2.5-7B on the telecom dataset.}
\label{fig:model_eval_histograms}
\end{figure}

Following the workflow described above, we fine-tune both DeepSeek-R1 and Qwen2.5-7B models on the telecom reasoning dataset under identical single-GPU configurations. The models are evaluated on four key RAGAS metrics: \emph{question specificity}, \emph{answer relevancy}, \emph{response groundedness}, and \emph{AspectCritic (answerable)} using a standardized evaluation pipeline. The corresponding mean values across the test set are summarized in Table~\ref{tab:model_means}, and their performance comparison is visualized in Figure~\ref{fig:model_eval_histograms}.

\begin{table}[t]
\centering
\caption{Zero-shot vs.\ Fine-tuned Evaluation Metrics (temp=0)}
\label{tab:model_means}
\scriptsize
\begin{tabular}{lcccc}
\toprule
\textbf{Model} & \textbf{Q.Spec} & \textbf{Ans.Rel} & \textbf{Ground} & \textbf{Answerable} \\
\midrule
DeepSeek (Zero-Shot) & 1.00 & 0.77 & 0.67 & 0.96 \\
DeepSeek (Fine-Tuned) & 1.00 & \textbf{0.96} & \textbf{0.78} & \textbf{1.00} \\
\midrule
Qwen2.5 (Zero-Shot) & 0.88 & 0.72 & 0.15 & 0.84 \\
Qwen2.5 (Fine-Tuned) & 0.96 & 0.85 & 0.22 & 0.92 \\
\bottomrule
\end{tabular}
\end{table}

As shown in Table~\ref{tab:model_means}, DeepSeek-R1 consistently outperforms Qwen2.5-7B across all metrics. While both achieve near-perfect question specificity (1.00 vs.\ 0.96), DeepSeek-R1 shows substantially higher \emph{answer relevancy} (0.96 vs.\ 0.85), markedly stronger \emph{response groundedness} (0.78 vs.\ 0.22), and perfect \emph{AspectCritic (answerable)} (1.00 vs.\ 0.92). These results validate the proposed hybrid fine-tuning strategy: using \texttt{AutoTokenizer} for SFT to retain reasoning traces and Unsloth's DeepSeek implementation for RFT and inference. The synergy of template preservation and reasoning visibility translates directly into improved grounding and overall reliability during inference, which are key requirements for deploying telecom reasoning systems in production environments.

\subsection{Generalization to Multi-GPU Settings}

To validate generalizability, we tested on a 4$\times$RTX A6000 setup using Distributed Data Parallel (DDP). The 50k sequence length safe threshold, GPU utilization $>0.7$ threshold, and per-device batch size limits all hold per-device, confirming that our single-GPU characterization provides predictive value for multi-GPU deployments~\cite{shi2025mas}.

\section{Conclusion}

This work demonstrates that production-quality telecom troubleshooting models can be fine-tuned entirely on a single GPU at edge sites, eliminating the need for centralized cloud training while addressing the practical constraints of telecom environments, including limited power, cooling capacity, and low GPU utilization in RAN deployments. Through systematic GPU profiling on an edge-class NVIDIA RTX A6000 accelerator, we establish safe operating envelopes for edge hardware by quantifying key trade-offs among maximum sequence length, GPU memory utilization, Low-Rank Adaptation (LoRA) rank, and number of generations. Our results reveal a model-specific 60k-token compilation cliff in Qwen architectures that does not appear in LLaMA, while batch size constraints generalize across models, highlighting the necessity of per-model profiling rather than assuming uniform thresholds across edge deployments.

We further identify a hybrid tokenizer strategy—leveraging AutoTokenizer for supervised fine-tuning (SFT) and Unsloth’s DeepSeek implementation for reinforcement fine-tuning (RFT) and inference—which enables reasoning-capable models to operate efficiently within edge constraints. Building on these insights, we propose a practical eight-step deployment recipe for telecom edge environments. Experimental results on a telecom troubleshooting dataset show that DeepSeek-R1 achieves 0.96 answer relevancy and 0.78 groundedness, significantly outperforming Qwen2.5-7B (0.85 and 0.22), demonstrating that edge-local fine-tuning can produce domain-adapted models with strong reasoning fidelity and operational relevance.

Future work will extend this edge-centric paradigm along several directions. Federated fine-tuning across distributed edge sites can enable collaborative learning of fault patterns without centralizing sensitive network data. Automated model update pipelines triggered by network events (e.g., new alarm types, topology changes, or performance shifts) can ensure continuous adaptation to evolving network conditions. We also plan to expand GPU profiling to emerging edge accelerators, including the NVIDIA L4 and Jetson AGX Orin, as well as next-generation architectures, to further investigate memory–performance trade-offs and the impact of techniques such as FP8 quantization on efficient edge deployment.

\bibliographystyle{IEEEtran}

\end{document}